# Thermal evolution of the early Moon

Sandeep Sahijpal and Vishal Goyal, Department of Physics, Panjab University, Chandigarh, India 160014 (sandeep@pu.ac.in)

**Abstract:**

The early thermal evolution of Moon has been numerically simulated to understand the magnitude of the impact induced heating and the initially stored thermal energy of the accreting Moonlets. The main objective of the present study is to understand the nature of processes leading to core-mantle differentiation and the production and cooling of the initial convective magma ocean. The accretion of Moon was commenced over a timescale of 100 years after the giant impact event around 30–100 million years in the early solar system. We studied the dependence of the planetary processes on the impact scenarios, the initial average temperature of the accreting moonlets and the size of the protoMoon that accreted rapidly beyond the Roche limit within the initial one year after the giant impact. The simulations indicate that the accreting Moonlets should have a minimum initial averaged temperature around 1600 K. The impacts would provide additional thermal energy. The initial thermal state of the moonlets depends upon the environment prevailing within the Roche limit that experienced episodes of extensive vaporization and re-condensation of silicates. The initial convective magma ocean of depth more than 1000 km is produced in the majority of simulations along with the global core-mantle differentiation in case the melt percolation of the molten metal through porous flow from bulk silicates was not the major mode of core-mantle differentiation. The possibility of shallow magma oceans cannot be ruled out in the presence of the porous flow. Our simulations indicate the core-mantle differentiation within the initial $10^2$-$10^3$ years of the Moon accretion. The majority of the convective magma ocean cooled down for crystallization within the initial $10^3$-$10^4$ years.

**Introduction**

The conventional hypothesis for the formation of Moon involves an oblique collision of a Mars-sized planet with the early Earth (Hartmann and Davis 1975; Stevenson 1987; Cameron 1997; Canup and Asphaug 2001; Canup 2004a, b, 2008, 2014; Asphaug 2014; Barr 2016). Based on the elemental composition and the chronological records of the Earth and the Apollo-returned lunar samples the giant impact is considered to have occurred within the initial 30 to > 100 million years (Ma) of the formation of the solar system (e.g., Touboul et al. 2007, 2015; Bottke et al. 2010; Borg et al. 2011; Yu and Jacobsen 2011; Elkins-Tanton et al. 2011; Kruijer et al. 2015). The Earth and, presumably the colliding planet, Theia, in the conventional giant impact scenario, are considered to have undergone substantial planetary scale differentiation into an iron core and silicate mantle prior to the collision (e.g., Melosh 1990; Canup 2004b; Salmon and Canup 2012, 2014). The collision of the two planetary bodies probably led to the merger of their iron cores with some extent of the mixing of the impactor's iron core with its molten silicate during the merger.

Several hypotheses have been proposed for the formation of Moon along with several subsequent refinements. These include, i) the fission hypothesis, ii) the cogenetic accretion of Moon along with Earth and, iii) the capture hypothesis (see e.g. reviews, Stevenson 1987; Canup 2004b; Asphaug et al. 2006; Barr 2016). The giant impact hypothesis is considered as the most viable scenario as it can explain some of the essential physical and chemical properties of Moon (Canup 2004b; Asphaug 2014; Taylor and Wieczorek 2014; Barr 2016). The high angular momentum of the Earth-Moon system compared to other planet-satellite(s) systems in the solar system supports the giant impact hypothesis. The substantial depletion of metallic iron in Moon as inferred from its low density (~3333 kg m$^{-3}$) indicates that the formation of the satellite could not be reconciled with the conventional understanding of the accretion of a planetary body in the solar nebula followed by planetary scale differentiation on account of accretion and impact induced energies (Lucey et al. 1995). Moon has an estimated small iron core of radius ~330 km (Weber et al. 2011; Williams et al. 2014) and a two billion years record of an early lunar dynamo (Tikoo et al. 2017). Further, the highly similar the isotopic composition of oxygen (Wiechert et al. 2001), chromium (Lugmair and Shukolyukov 1998), silicon, titanium (Zhang et al. 2012) and tungsten (Touboul et al. 2007) among Moon and Earth indicates the possibility of a substantial lunar mass derived from the Earth at the time of giant impact (Meier et al. 2014; Mastrobuono-Battisti et al. 2015). Finally, the low abundance of the volatiles, K, Na and Zn (Wolf and Anders 1980; Taylor et al. 2006) indicates the formation of Moon by a cataclysmic event that resulted in substantial volatile depletion probably due to incomplete accretion by Moon (Canup et al. 2015).

The giant impact collision resulted in the formation of a protolunar debris disk around the Earth. Moon is considered to have formed from the rapid accretion of this disk. A wide range of numerical simulations have been performed to understand the dynamics of the collision, the formation and evolution of the disk (see e.g., Canup and Esposito 1996; Machida and Abe 2004; Wada et al. 2006; Canup et al. 2013; Charnoz and Michaut 2015; Hosono et al. 2016). In the earliest versions of the conventional giant impact scenario, a disk with twice the mass of Moon was formed (e.g., Canup 2008). The rapid accretion of Moon could have occurred over a timescale of ~1 year with most of the material acquired from the impactor (Ida et al. 1997; Canup 2004a). The energy generated by the rapid accretion of Moon could have resulted in the global scale melting of Moon. Further, the rapid accretion could have hindered the equilibration of the accreting Moon material with the Earth's mantle ejected debris field that is essential to acquire the observed isotopic similarity between Earth and Moon.

The rapid accretion of Moon was circumvented by partitioning of the disk material due to the Roche limit at a distance of ~3 Earth's radii. As the protolunar disk evolved, the matter beyond the Roche limit could have accreted massive clumps (moonlets) due to gravitational instabilities leading to the rapid formation of protoMoon. However, the clumps of the accreted matter within the Roche limit were physically churned by the strong gravitational tides of Earth and resulted in substantial collisions, thereby, leading to the heating and vaporization of silicates in the disk. The re-condensation of silicate vapors occurred during the cooling of the disk within the Roche limit over a timescale of 100 years (Thompson and Stevenson 1988; Salmon and Canup 2012, 2014). This increased the possibility of equilibration of the protolunar disk material with the Earth's mantle generated debris so that the Moon acquired an isotopic composition identical to Earth's mantle (Pahlevan 2014; Pahlevan and Stevenson 2007; Pahlevan et al. 2011; Zahnle et al. 2015). The simulations based on three-stage accretion of Moon from a protolunar disk over a timescale of 100 years have been performed with a protracted contribution of the accreting matter from the inner regions of the Roche limit to the protoMoon undergoing formation beyond the Roche limit (Salmon and Canup 2012, 2014). One of the major advantages of this scenario is the isotopic homogenization acquired by the matter within the Roche limit with the Earth's mantle debris field.

In the conventional (canonical) giant impact scenario more than 70% of the initial mass of the protolunar disk is derived from the impactor's mass (Canup 2004a), thereby making it difficult to reconcile the isotopic similarity between Earth and Moon. The substantial contribution of Earth's mass to the protolunar disk would require large impactor energy. Apart from the conventional giant impact scenarios, the recent non-canonical hypothesis for the formation of Moon with large excess of angular momentum has been proposed based upon the collision between two planetary embryos during the final stages in the evolution of planetary system either involving a massive impactor or a high velocity impactor undergoing a head-on collision with fast spinning Earth (Canup 2012; Ćuk and Stewart 2012). In addition, the possibility of a hit and run scenario has been proposed (Reufer et al. 2012). The generated non-canonical disks are more compact compared to the canonical disks. These disks have higher initial mass fraction within the Roche limit. The numerical simulations based on three-stage accretion scenario for the formation of Moon yield almost identical results for the canonical and non-canonical accretion scenarios (Salmon and Canup 2012, 2014). Thus, irrespective of whether the giant collision involved the canonical or the non-canonical collisional scenario, the accretion of Moon probably occurred over a time scale of hundreds of years, with the majority of the accretion within the initial 100 years.

Apart from the significant depletion of metallic iron that is associated with a small iron core of radius ~330 km (Weber et al. 2011; Williams et al. 2014), Moon has a thick anorthositic-rich crust with an estimated thickness of 34-43 km (Wieczorek et al. 2013). The thickness of the crust depends upon the depth of the initial magma ocean generated at the time of the formation of Moon (Warren 1985). A depth of 200-300 km has been estimated for the initial magma ocean (e.g., Barr 2016). However, a magma ocean with a depth up to 1000 km has been also proposed that is consistent with an anorthositic crust of 40-50 km (Elkins-Tanton et al. 2011). This magma ocean is considered to have substantially cooled until solidification within a timescale of ~1000 years. This imposes stringent constraints on the thermal models dealing with the early thermal evolution of Moon. The three-stage accretion scenario for Moon (Salmon and Canup 2012, 2014) results in a protracted accretion over a timescale of several hundreds of years. This scenario should result in a gradual rise in the temperature of the accreting Moon on account of impact energy to produce a partially molten magma ocean. In the present work, we have performed detailed numerical simulations of the accretion and the early thermal evolution of Moon to understand the extent of melting, the dynamics of the core-mantle differentiation and the cooling of the convective magma ocean.

Attempts have been made previously to understand the early thermal models of Moon (e.g., Kaula 1979; Pritchard and Stevenson 2000; Elkins-Tanton et al. 2011; Barr 2016) by parametrically incorporating the impact generated heat during accretion. Although the models are capable of adequately heating Moon by impacts, these models cannot impose physical constraints on the efficiency with which the impact energy is stored at the surface (Schubert et al. 1986). One of the objectives of the present work is to impose constraints on the initial average temperature of the accreting Moonlets and the efficiency of impact heat generation by numerically simulating the early thermal evolution of Moon, starting from its accretion to planetary scale differentiation leading to the formation of an iron core and a convective magma ocean.

**Thermal models and numerical simulations**

The protracted accretion scenario for the accretion of Moon from the protolunar disk over a timescale of ~100 years (Salmon and Canup 2012, 2014) forms the basis of the thermal models developed in the present work. The thermal evolution of Moon was numerically executed by solving the heat conduction partial differential equations (Equation 1) that incorporate the contribution of the long-lived radionuclides, $^{40}$K, $^{235}$U, $^{238}$U, and $^{232}$Th. As discussed earlier, the formation of Moon could have occurred during the initial 30 to > 100 million years (Ma) of the formation of the solar system (e.g., Touboul et al. 2007, 2015; Yu and Jacobsen 2011; Kruijer et al. 2015). We assumed the formation time to be 50 Ma. However, this time could be temporally translated within the range of 30 to 100 Ma as the activities associated with the long-lived nuclides remain relatively unaltered during these timescales. The contribution of heat from the short-lived nuclides, $^{26}$Al and $^{60}$Fe (Sahijpal et al. 2007), is not explicitly relevant for the thermal evolution of Moon. However, the role of these short-lived nuclides in the thermal evolution of planets and their planetary embryos is certainly relevant (Sahijpal and Bhatia 2015; Bhatia and Sahijpal 2016, 2017a, b), and could be of relevance even for the early heating of Earth and the impactor responsible for the formation of Moon. Hence, the implicit contribution of short-lived radioactive heating at least in Earth and Theia cannot be rejected. Nonetheless, we ignore the direct contribution of the short-lived nuclide heating in the thermal models of Moon.

The heat conduction differential equation (Equation 1) was numerically solved by using the finite difference equation (Sahijpal et al. 1997; Gupta and Sahijpal 2010; Sahijpal and Bhatia 2015; Bhatia and Sahijpal 2016, 2017a, 2017b). The final radius and the mass of Moon are assumed to be 1740 km and $7.35 \times 10^{22}$ kg, respectively.

$$\frac{\partial T}{\partial t} = \kappa \ \nabla^2 T + Q(t) \qquad (1)$$

Here, $T(r, \theta, \varphi, t)$ is a function of spherical polar coordinates, $r, \theta, \varphi$, and time, 't'. $Q(t)$ is the rise in the temperature on account of the contribution of heat from long-lived nuclides, and '$\kappa$' is the temperature dependent thermal diffusivity. Equation 1 was converted into a one-dimensional spatial differential equation involving the radial coordinate, 'r' (Bhatia and Sahijpal 2017a) by assuming Moon to be spherically symmetric during and after its accretion. The one-dimensional equation was converted into a finite difference equation using classical explicit approximation (Lapidus and Pinder 1982). The simulation parameters for the various thermal models are mentioned in Table 1. The time, '$T_{Onset}$', represents the onset time of the accretion of Moon that marks the time of giant impact. The time, '$T_{acc.}$', represents the accretion time of Moon. Equations 2-4 represent the numerical solutions and the surface condition of the partial differential equation.

$$T_{0,j+1} = T_{0,j}(1 - 6\kappa \ \sigma) + T_{1,j} 6 \kappa \ \sigma + Q(t) \ \delta t \qquad (2)$$

$$T_{i,j+1} = T_{i-1,j}\left(1 - \frac{1}{i}\right) \kappa \ \sigma + T_{i,j}(1 - 2 \kappa \ \sigma) + T_{i+1,j}\left(1 + \frac{1}{i}\right) \kappa \ \sigma + Q(t) \ \delta t \qquad (3)$$

$$T_{S, j+1} = 250 \ K \qquad (4)$$

Here, $\sigma = \frac{\delta t}{(\delta r)^2}$, where $\delta t$ and $\delta r$, are the fixed sizes of the temporal and spatial grids, respectively. Most of the simulations were performed by assuming the temporal and spatial grid sizes of 0.1 years (36.5 days) and 2 km, respectively, except in the case of one simulation corresponding to the Model J where the temporal grid size was assumed to be 0.01 years (3.65 days). Equation 3 numerically predicts the temperature, $T_{i,j+1}$, at the spatial and temporal grids, 'i' and 'j+1', respectively, on the basis of the temperature at an earlier temporal grid, 'j' at the spatial grids, 'i-1', 'i' and 'i+1'. While Equation 2 represents the thermal evolution at the central spatial grid '0', the Equation 4 represents the fixed surface temperature condition at the defined surface of Moon during and subsequent to its accretion. The thermodynamical aspects related with the temperature dependent specific heat and thermal diffusivity, the melting of the planetary body and the segregation of metallic iron from silicate melt to form an iron core and silicate mantle were numerically executed to understand the thermal evolution.

**The initial condition, composition and the accretion scenarios of Moon**

The accretion scenario adopted in the present work is based on the anticipated three-stage accretion of Moon from the protolunar disk (Salmon and Canup 2012, 2014). Variations in the accretion scenarios have been considered to explore the parametric space associated with the numerical simulations. The accretion scenario parameters and the assumed initial temperatures for the various thermal models are presented in Table 1. Subsequent to the giant impact, we initiated the majority of the simulations with the formation of a protoMoon within a time-span of one year by the accretion of matter beyond the Roche limit where the matter does not suffer substantial gravitational tidal influence of the Earth. It should be mentioned that the majority of the simulations have a temporal resolution of 0.1 year. The accretion of the protoMoon with two distinct assumed seed sizes (Table 1) was considered to study the influence of the accretion scenario on the thermal evolution. For example, in the case of Models A-F, and Models G-L, the radius of the seed Moon was assumed to be 1074 and 1240 km, respectively, that corresponds to 0.24 and 0.36 mass fraction of the lunar mass accreting during the first year beyond the Roche limit (Salmon and Canup 2014).

The accretion of the remaining mass fraction of Moon was numerically commenced during 100 years (Salmon and Canup 2014). As mentioned earlier, the protolunar disk material within the Roche limit experienced substantial silicate vaporization followed by isotopic equilibration with the Earth's mantle ejected debris field. Finally, this vaporized material re-condensed and a fraction of this material eventually drifted out beyond the Roche limit and got accreted on protoMoon over a timescale of ~100 years. We have considered a linear accretion, $'r(t) \propto t^n, n = 1'$, of Moon in terms of its radial size (Merk et al. 2002; Sahijpal et al. 2007; Bhatia and Sahijpal 2017a). Hence, the lunar mass, 'M(t)', will increase according to the relation $M(t) \propto t^3$. The accretion was numerically executed by systematically adding uniform sized spatial grids to an already existing array of spatial grids defining Moon at specific time intervals till Moon acquires its final size (see e.g., Sahijpal et al. 2007).

The Moon was assumed to have accreted uniformly hot at high temperature during its accretion. The initial temperature at the onset of the accretion of Moon was assumed in the range of 1400-1700 K depending upon the model (Table 1). The simulations performed with an initial temperature lower than this range results in an undifferentiated Moon. The numerical simulations based on Equation 1 were initiated by assuming an initial temperature for all the spatial grids except for the surface of Moon during and subsequent to its accretion that was defined according to Equation 4. The additional rise in the temperature at various spatial grids due to impact induced heating (Equation 5) was further incorporated in the finite difference heat conduction equations. The range of 1400-1700 K for the initial temperature was considered in order to study the role of the thermal state of the accreting matter prevailing in the protolunar disk that had suffered episodes of extensive heating, vaporization and cooling (Pritchard and Stevenson 2000). Moonlets formed by the accretion of the re-condensed silicate grains can maintain high temperatures against conductive heat losses from their surface over a time-span of 100 years during which Moon accretes majority of its mass.

The low density (~3333 kg m$^{-3}$) of Moon indicates substantial depletion of metallic iron and a small-sized iron core of radius ~330 km (Weber et al. 2011; Williams et al. 2014). The densities of iron core and silicate mantle subsequent to differentiation were assumed to be 5251 kg m$^{-3}$ and 3300 kg m$^{-3}$. The initial metallic iron, nickel and FeS compositions of Moon were assumed to be 0.0288 times the corresponding H-chondrite abundances (Jarosewich 1990; Sahijpal et al. 2007). This implies an initial metallic iron, iron in FeS, sulfur and nickel abundances of 0.634, 0.247, 0.142 and 0.052 wt. %, respectively. This constitutes a total mass percentage of ~1.07 wt. % for Fe-Ni-FeS that eventually forms an iron core of radius ~328 km. The bulk silicates constitute the remaining mass. Further, Moon has an estimated 12-14 wt. % of FeO (Sossi and Moynier 2017).

**The impact energy associated with the accretion of Moon**

The impact energy generated during the accretion of Moon was incorporated by following the criteria developed in the earlier works (Kaula 1979; Schubert et al. 1986; Bhatia and Sahijpal 2016, 2017a, b). The energy generated during impacts is converted into an associated rise in the temperature (Equation 5) above the prevailing temperature of a specific spatial grid defined by the finite difference method. During the accretion of the planetary body, the temperature at the transitory region near the surface spatial grid of the body at a specific time is augmented by the temperature estimated by Equation 5. Here, 'M(r)' is the mass of the body at a specific instant of accretion, 'c' is the specific heat, 'u' is the relative approach velocity per unit mass of the incoming Moonlet. The parameter, 'h' is an uncertain efficiency parameter which determines the extent of impact energy that is effectively translated in terms of the temperature rise against heat losses from the surface (Kaula 1979; Pritchard and Stevenson 2000; Barr 2016; Bhatia and Sahijpal 2016, 2017a, b). The earlier works based on the thermal models of planetary bodies suggest that the contribution of the second term in the parentheses of Equation 5 is negligible (Bhatia and Sahijpal 2017a), and the efficiency parameter, 'h' has been generally considered to be of the order of 0.1. In order to accrete Moon against disruptive collision induced breakdowns, the relative approach velocity per unit mass, 'u', of the incoming Moonlets should be always less the escape velocity, $\{2GM(r)/r\}^{1/2}$, in the range of <1 to 2.38 km s$^{-1}$, during the distinct accretionary stages. The negligible contribution from the second term in the Equation 5 would ensure this condition. We have also not considered large impacts that could result in large scale melting. The role of the h parameter has been extensively studied in the present work. Further, as elaborately discussed in the following, we have made use of a modified relaxation method developed by Senshu et al. (2002) on account of high convective heat losses to generate a thermal gradient inside Moon according to the gradient in the melting temperature of the bulk silicate. This results in less prolonged heating, specifically in the outer regions, on account of high convective heat losses from the body in comparison to that estimated by Equation 5.

$$T(r) = h \frac{GM(r)}{c\,r} \left(1 + \frac{r u^2}{2GM(r)}\right) \qquad (5)$$

**The role of the long-lived nuclides in the thermal evolution**

The long-lived nuclides, $^{40}$K, $^{235}$U, $^{238}$U and $^{232}$Th, provide thermal energy over a timescale of a billion years. The major contributions of the long-lived nuclides in the thermal evolution of icy bodies in the early solar system have been recently inferred (Bhatia and Sahijpal 2017b). The long-lived nuclides could have provided substantial heat to trigger even core-mantle differentiation in icy bodies. The initial estimates of these nuclides in the early solar system (Lodders et al. 2009) indicate maximum thermal contribution (~80%) from $^{40}$K (Bhatia and Sahijpal 2017a). However, the estimated abundance of potassium ($^{39}$K) is quite low in the case of volatile-depleted Moon (0.0037 wt. %; Taylor and Wieczorek 2014) compared to the rocky planetary bodies (0.07 wt. %) that formed from the accretion of solar nebula condensates (Table 1). Due to the depletion of potassium in Moon, the long-lived nuclides were not able to appreciably raise the interior temperature to cause core-mantle differentiation. Hence, the thermal contributions of these nuclides seem to be substantially less for the global scale heating of Moon. This is distinct from the thermal evolution of icy bodies that indicate significant thermal contributions from long-lived nuclides over the initial 1-2 billion years (Bhatia and Sahijpal 2017b). However, the concentration of these radionuclides within the crust after the large-scale planetary differentiation of Moon would result in radiogenic heating of the crust.

**Thermodynamics of the thermal models**

The temperature dependent thermal diffusivity ($\kappa$) of the un-melted Moon was assumed in the range of $(6.4-5.4) \times 10^{-7}$ m$^2$ s$^{-1}$ during the thermal evolution, whereas, the specific heat of the un-melted planetary body was assumed to vary within, 610–830 J kg$^{-1}$ K$^{-1}$, in the temperature range 250–1450 K (Sahijpal et al. 2007). The impact induced heating during the accretion along with the initial average temperature of the accreting Moonlets could eventually lead to a gradual rise in the temperature of Moon far beyond the melting temperatures of metallic iron and silicate mass fractions. We have not incorporated the gravitational tidal energy in our simulations that could substantially influence the outer cooling regions (Meyer et al. 2010). The tidal energy could also substantially heat the accreting Moon, specifically during the first year. The solidus and the liquidus temperatures of the metallic Fe-FeS were assumed to be 1213 K and 1233 K, respectively, at 0 Pa pressure. The solidus and the liquidus temperatures of the silicate were assumed to be 1450 K and 1850 K, respectively, at 0 Pa pressure (Taylor et al. 1993). The values of the solidus and liquidus temperatures were appropriately scaled on account of the distinct hydrostatic pressures prevailing at different depths within Moon. The details of the procedure can be found in literature (Senshu et al 2002; Sahijpal and Bhatia 2015; Bhatia and Sahijpal 2016, 2017a). These inferred phase transition temperatures are represented in Figure 1 at distinct depths along with the deduced pressure-depth profile in the fully accreted undifferentiated Moon with an assumed uniform density. The variation of the pressure-depth dependence was considered during the accretion of Moon along with its influence on the solidus and liquidus temperatures of silicate and metallic Fe-FeS. Once a spatial region within Moon acquires the melting temperature for either metallic iron or silicate, the latent heat of phase transition was incorporated in the specific heat by following the procedure adopted by Merk et al. (2002) and Sahijpal et al. (2007). The latent heat of Fe-FeS melting and silicate melting were assumed to be $2.7 \times 10^5$ J kg$^{-1}$ and $4.0 \times 10^5$ J kg$^{-1}$, respectively (see e.g., Sahijpal et al. 2007). The specific heats of silicate and Fe-FeS melts were assumed to be 2000 J kg$^{-1}$ K$^{-1}$. The substantial melting of Moon results in the formation of a convective magma ocean whose thermal evolution is modelled by raising the thermal diffusivity in Equations 2 and 3 by several orders of magnitude in order to imitate the rapid cooling of the convective magma ocean (Sahijpal et al. 2007; Gupta and Sahijpal 2010; Neumann et al. 2014; Bhatia and Sahijpal 2017a).

**Convection in the magma ocean**

Due to our choice of the classical explicit approximation (Lapidus and Pinder 1982) in numerically solving the heat conduction partial differential equation, the simulations with several orders of magnitude high thermal diffusivity, referred as an *effective thermal diffusivity*, can be run with numerical stability only with a small temporal grid size of less than 0.1 years or 0.01 years in order to rapidly cool down the convective magma ocean. The simulations with the latter choice of temporal grid size take approximately one week of computing time to simulate the initial one million years of the thermal evolution of Moon. A further reduction in the temporal grid size would make it practically difficult to run a single simulation. In order to resolve the problem, we have adopted a modified form of the relaxation method proposed by Senshu et al. (2002) during the accretion growth of Moon that results in the thermal relaxation of the body against the high thermal energy deposited by impact collisions received during the accretion. Further, subsequent to the complete accretion of Moon, we have developed an approach of deciphering the nature of the cooling rate of the lunar convective magma ocean based on the choice of distinct effective thermal diffusivity values that spans over several orders of magnitude. Based on this analysis, we decipher the cooling rate for an appropriate high effective thermal diffusivity values that are estimated theoretically on the basis of detailed studies of the cooling convective magma ocean. In the following, we estimate the theoretical range of effective thermal diffusivity values to begin with, and then we proceed to decipher the nature of the cooling convective magma ocean for these theoretical effective thermal diffusivity values.

In order to model the heat transport due to convection, we followed the method presented by Solomatov (2007) to calculate the effective thermal diffusivity, '$\kappa_{conv.}$', which imitates the heat transport due to convection in magma ocean. This method has also been used for modeling of local magma ponds on Mars (Golabek et al. 2011), and for modeling of a shallow magma ocean on Vesta (Neumann et al. 2014). The convection has been computed in the soft turbulence regime where the Rayleigh number (Equation 6) is below approximately $10^{19}$, and in the hard turbulence regime where the Rayleigh number is above $10^{19}$. We have neglected the effects of rotation of Moon on its axis.

$$Ra_i = \frac{\alpha_i g_i (T_i - T_s) \rho_i L_i^3}{\kappa_i \eta_i} \qquad (6)$$

Here, for a specific spatial grid interval, 'i', $\kappa_i$ is thermal diffusivity, $g_i$ is the acceleration due to gravity, $L_i$ is the depth of magma ocean. We assume the viscosity ($\eta_i$), thermal expansivity ($\alpha_i$), density ($\rho_i$) and the surface temperature of magma ocean ($T_s$) to be constant for calculating the Rayleigh number. These values were assumed to be 1 Pa s, $2\times10^{-5}$ K$^{-1}$, 3333 kg m$^{-3}$ and 1625 K, respectively. With these assumptions, we get,

$$Ra_i = A_1 \, f_1(T) \, g_1(r/R) \qquad (7)$$

$$A_1 = \frac{4\pi G \alpha \rho^2 R^4}{3\eta}, \quad f_1(T) = \frac{T - T_s}{\kappa}, \quad g_1(r/R) = \frac{r}{R}\left(1 - \frac{r}{R}\right)^3 \qquad (8)$$

Here, 'r' represents the radial distance from the center of Moon where the various parameters have been evaluated and 'R' is the radius of Moon.

Melt fraction dependent Rayleigh Number

The above discussed Rayleigh number does not incorporate the effects of partially molten material. However, it may be seen that the only factor which heavily depends upon melt fraction is viscosity. Melt fraction $\phi$ dependent viscosity at high strain rates can be calculated according to Golabek et al. (2011):

$$F_\eta = \frac{\eta_l}{\eta} = \exp\left\{\left[2.5 + \left(\frac{1-\phi}{\phi}\right)^{0.48}\right](\phi - 1)\right\} \quad (9)$$

$$Ra(\eta) = Ra(\eta_l) F_\eta \quad (10)$$

Melt fraction $\phi$ is assumed to be linear with temperature between solidus and liquidus, with zero at temperature below solidus and one at temperatures above liquidus.

$$\phi = \frac{T - T_{sol}}{T_{liq} - T_{sol}} \quad (11)$$

In the Soft turbulence regime ($1418 < Ra_i < 10^{19}$)

$$\kappa_{conv.} = 0.089 \, Ra_i^{1/3} \kappa \quad (12)$$

In the Hard turbulence regime ($Ra_i > 10^{19}$)

$$\kappa_{conv.} = 0.22 \, Ra_i^{2/7} \, Pr^{-1/7} \, \lambda^{-3/7} \kappa, \quad Pr = \frac{\eta}{\rho\kappa} = \frac{\eta_l}{\rho\kappa F_\eta}, \quad \lambda = \frac{1}{\pi}\left(\frac{1 - r/R}{1 + r/R}\right) \quad (13)$$

Where $Pr$ is Prandtl number and $\lambda$ is the aspect ratio for the mean flow. Solomatov (2007) has argued that $\lambda = 1$ would be the simplest assumption because of the uncertainties associated with spherical symmetry. However, the effects of using the above expression, or $\lambda = 1$ have been only minor (Solomatov 2007). Using all the above expressions, we found $\kappa_{conv.}$ to rise up to a value of $\sim 5 \times 10^6 \, \kappa_{cond.}$ in the temperature range of 1625-2000 K.

Because it is not possible for us to numerically simulate the convective magma oceans at these high effective thermal diffusivities with numerical stability, we have separately considered the thermal evolution of Moon during and subsequent to its accretion by following two distinct approaches. We followed a modified form of the relaxation method identical to the one developed by Senshu et al (2002) to thermally relax the Moon against the impact induced thermal energies deposited during its accretion. During the accretion of Moon, as the temperature increases on account of deposition of the impact induced energy according to Equation 5, the thermal gradient across the body is established according to the gradient of the assumed melting temperature corresponding to ~43 % bulk silicate melting. This would correspond to a temperature of 1625 K near the surface spatial grid at 0 Pa pressure. This temperature is appropriately scaled in the inner region according to the pressure-depth melting temperature relation corresponding to the assumed ~43 % bulk silicate melting. The specific choice of the near-surface temperature of the magma ocean at 1625 K is made to ensure substantial convective heat losses from the magma ocean that is otherwise not achievable with our presently adopted effective thermal diffusivity. Further, the gravitational tidal interaction could provide substantial energy to the surface. We have made an assumption that the fully convective magma ocean is established within 40-50 % bulk silicate melting. The highly convective magma ocean rapidly cools down almost instantaneously during the accretional timescales of ~100 years. Thus, in our simulations, as the body accretes mass, the impact energy (Equation 5) effectively heats up the specific spatial region associated with the accretion at a specific time till the region acquires the melting temperature corresponding to ~43% bulk silicate melting. Thus, the modified relaxation method results in the creation of a thermal gradient across the convective magma ocean within the planetary body according to the bulk silicate melting temperature. This gradient serves as the initial condition for the subsequent cooling of the magma ocean as discussed in the following. It should be mentioned that if we do not adopt the modified relaxation method we will have to cool the initially generated high-temperature magma ocean by using high effective thermal diffusivity as indicated by the theoretical analysis. Since, we cannot perform simulation with this high effective thermal diffusivity, we end up with a sharp rise in the temperature near the surface by using the maximum effective thermal diffusivity as permissible with our numerical approach. This produces an inverted thermal gradient against convective heat flow. The inverted thermal gradient will prolong the further convective cooling of the interior of the magma ocean till the gradient subsides with time.

The subsequent cooling of the magma ocean after the complete accretion of Moon and thermal relaxation was numerically executed by considering a parametric Sigmoidal function for the effective thermal diffusivity according to Equation 14. In order to achieve numerical stability without any discontinuity, the Sigmoidal function ensures a gradual logarithmic reduction of the effective thermal diffusivity from an assumed defined maximum value of $\kappa_{conv.max.}$ at 1650 K to $\kappa_{cond.}$ at 1550 K at 0 Pa pressure. These two temperatures correspond to 50 and 25 % of bulk silicate melting, respectively. We assume that the convection in the magma ocean substantially reduces as the bulk silicate abundance in the melt form become less than 25 % on account of crystallization and Ostwald ripening (Solomatov 2007). At 1625 K, the $\kappa_{conv.}$ acquires half of the maximum effective thermal diffusivity value, '$\kappa_{conv.max.}$'. Appropriate scaling to the above-mentioned specific temperatures in the range of 1650 -1550 K in the inner regions, according to the pressure-depth melting temperature relation, were performed by defining the temperature, '$T_{Ref.}$' (Equation 14). This reference temperature was assumed to be 1625 K at 0 Pa pressure near the surface and was appropriately scaled inwards.

Neumann et al. (2014) have numerically simulated the cooling of the magma ocean on Vesta by considering the dependence of the effective thermal diffusivity at a spatial grid interval, 'i' on the prevailing temperature, '$T_i$', the acceleration due to gravity, 'g(r)' and the depth of the magma ocean, '$L_i$'. The majority of the remaining parameters (see e.g., Equation 6) do not exhibit major variations. We could incorporate the variation due to the acceleration due to gravity in a normalized manner in the effective thermal diffusivity (Equation 14, below) by defining the acceleration due to gravity at the lunar surface, 'R', as, g(R), and a specific distance, 'r', from the center, as g(r). We could not successfully incorporate the influence of the depth of the magma ocean as its incorporation results in an increase in the temperature near the surface in our simulations that eventually leads to the inverted thermal gradient near the surface. This is one of the shortcomings of the present approach.

$$\kappa_{conv.} = \left(\frac{g(r)}{g(R)}\right)^{0.33} \times \frac{\kappa_{conv.max.}}{1+e^{-\frac{T_i-T_{Ref.}}{7.0}}} \qquad (14)$$

It should be noted that the cooling of the magma ocean in the case of Moon is distinct from that in the case of Vesta due to two major reasons. Firstly, the impacts deposit more energy in the exterior regions of Moon during the accretionary phase, whereas in the case of Vesta the heating is essentially due to short-lived radionuclides. Secondly, due to the higher interior hydrostatic pressures inside Moon compared to Vesta, the melting (and solidification) temperatures of silicates significantly change within the body. These two differences could significantly influence the cooling rates of the two planetary bodies.

The effective thermal diffusivity was incorporated in the numerical solutions of the heat conduction partial differential equation in a manner so that the conductive part of the heat flow takes place both inwards as well as outwards, whereas, the convective part of the heat flow takes place only in the outward direction. This is numerically executed by dividing the net thermal diffusivity, as mentioned in Equation 3, as a sum of the heat conductive diffusivity and the effective thermal diffusivity representing the convection. This ensures that the inner regions of the magma ocean cool down by an efficient heat flow through the outer regions and surface. Hence, the thermal gradient in the inner regions gradually declines as a result of cooling and solidification. The extent of the depth of the effective convective magma ocean from the surface reduce gradually over time. The value of '$\kappa_{conv.}$', according to Equation 14, in the outer most spatial grid was estimated by assuming a $T_{Ref.}$ temperature of 1625 K in order to ensure maximum outflow of the thermal energy from the surface for rapid cooling of the magma ocean. This assumption will hold in majority of the simulations that were run for the initial couple of million years. We have not considered the formation of an anorthositic crust that would substantially hinder the out-flow of the thermal energy. The continuous break down of the anorthositic crust by the surface impacts would maintain rapid outward convective heat flow (Perera et al. 2017).

As mentioned earlier, the temporal grid size in majority of simulations in the present work was chosen to be 0.1 years (36.5 days) with a spatial grid size of 2 km. This imposes a constraint on the value of $\kappa_{conv.max.} / \kappa_{cond.}$ (Equation 14) within the framework of the presently adopted finite difference method. We successfully ran the majority of the simulations with a value of $2\times10^4$ for $\kappa_{conv.max.} / \kappa_{cond.}$ without any numerical instability. The numerical instabilities in the corresponding simulations appear as we increase the value to $\sim10^5$ for $\kappa_{conv.max.} / \kappa_{cond.}$. For a specific choice of the simulation parameters dealing with '$T_{ini.}$' and 'h', we ran two additional simulations with two distinct values of $2\times10^3$ and $4\times10^4$ for $\kappa_{conv.max.} / \kappa_{cond.}$ in order to understand the dependence of the cooling rate of the convective magma ocean on the effective thermal diffusivity. We also ran a simulation with the temporal grid size of 0.01 years (3.65 days) that enabled us to increase the value to $2\times10^5$ for $\kappa_{conv.max.} / \kappa_{cond.}$. We observed a well-defined behavior on the basis of the simulations with four distinct values of $2\times10^3$, $2\times10^4$, $4\times10^4$ and $2\times10^5$ for $\kappa_{conv.max.} / \kappa_{cond.}$. The cooling timescales of the convective magma ocean systematically reduce with the increase in the value of $\kappa_{conv.max.} / \kappa_{cond.}$. On the basis of this inference, we extrapolated the evolutionary trends for the cooling rates of two distinct higher values of $2\times10^6$ and $2\times10^7$ for $\kappa_{conv.max.}$ / $\kappa_{cond.}$ in case of all simulations, thereby, predicting the thermal evolution of the cooling magma ocean in a manner that is compatible with the theoretical predictions. The extrapolation at these two higher values of the effective thermal diffusivities would define a probable extreme bound on the range of the effective thermal diffusivity. Thus, the modified relaxation method adopted for rapidly cooling of the initial convective magma ocean during the accretional phase of Moon, followed by the parametric approach for the gradual cooling of the magma ocean would provide a general overview of the early thermal evolution and planetary scale differentiation of Moon. Finally, we anticipate that the approach adopted in the present work is not critically sensitive to the small variations in the assumed parameters, specifically related with Equation 14. Further, we do not expect significant changes in case the relaxation method is applied at 50 % bulk silicate melting instead of the presently adopted 43 % bulk silicate melting requirement.

Subsequent to substantial heating and melting of Moon, the core-mantle differentiation would be initiated. Rushmer et al (2000) suggested that the metallic melts generated during melting do not segregate under static conditions unless melt percentage of silicates is high at a level of 40 % in the absence of high contents of volatiles in metallic melt. This is due to the high interfacial energies between molten metal and solid silicate, and the dihedral angle being much greater than 60°. In such a scenario it is difficult for metal melt to move through solid silicate matrix or form interconnect channels/veins. Even if the metallic melt was rich in volatiles, or there were deformations caused by dynamic conditions, and the segregation have started at lower melt fraction, Moon was volatile depleted. Rushmer et al. (2000) have argued that metal droplets would sink to the base of a 700-800 km deep magma ocean in only 10-100 years. The planetary scale core-mantle differentiation of Moon was numerically executed in a manner identical to the criteria adopted by Sahijpal et al. (2007). The segregation of metallic melt from silicate mush was triggered subsequent to 40 % bulk silicate melting in the majority of our simulations (Taylor et al. 1993; Rushmer et al. 2000). The metallic melt generated within the spatial grid interval was moved towards the center of Moon according to the viscosity dependent descent velocity of the metallic melt blobs through the silicate mush (Sahijpal and Bhatia 2015). In the present work, we present generalized results based on an order of magnitude variations in the viscosity dependent descent velocity of the metallic melt blobs. The accumulation of the metallic melt blobs at the center pushes outwards the silicate mush due to buoyancy. The numerical details of the differentiation criteria can be found in recent work (Sahijpal and Bhatia 2015; Bhatia and Sahijpal 2016, 2017a). Even though most of the simulations were performed with the requirement of 40 % bulk silicate melting to initiate core-mantle differentiation, we performed a simulation with 10 % bulk silicate melting to trigger differentiation. We treat the silicate melt fraction as one of the simulation parameters. Finally, we have also presented the timescale over which the core-mantle differentiation could occur by melt percolation, following the Darcy's law.

**Results and Discussion**

A representative set of a dozen simulations were performed with the basic aim to understand the early thermal evolution and the core-mantle differentiation of Moon (Table 1). The extent of the depth of the initial magma ocean generated by substantial silicate melting was also studied as the depth of this initial magma ocean would determine the thickness of the anorthositic-rich crust. Based on the present work, we find a strong correlation between the extent of the core-mantle differentiation and the depth of the initial magma ocean. The simulations were performed with distinct values of the initial temperature ($T_{ini.}$), the initial seed radius and the impact energy efficiency parameter, 'h' to explore wide possibilities of the accretion scenarios (Table 1). The initial accreting temperature ($T_{ini.}$) represents the initial average temperature of the accreting Moonlets on Moon. We have not considered the gravitational tidal energy in our simulations (Meyer et al. 2010). Based on our experience gathered during the development of the numerical code we focus our discussion on a narrow range of viable simulation parameters that are critical to understand the plausible early thermal evolution of Moon. The value of the accreting temperature ($T_{ini.}$) is varied over the range of 1400-1700 K that is sufficient enough to impose constraints on the initial average temperature required to trigger core-mantle differentiation in Moon. The choice of any initial average temperature lower than this range appears to be formidable, whereas, a higher choice will definitely result in the core-mantle differentiation with a comparatively large initial convective magma ocean. The impact induced efficiency parameter, 'h' in most of the simulations is confined within the range of 0.1-0.2 in order to be consistent with the thermal models of other planetary bodies (see e.g., Bhatia and Sahijpal 2016, 2017a, b) that indicate lower values. However, we ran several simulations with a value of 0.5 for h. The radius of the seed protoMoon that accreted rapidly within the first year (Salmon and Canup 2014) was assumed to be either 1074 or 1240 km in the simulations. We have not considered major variations in this parameter as we could not find any major dependency of the thermal evolution on this parameter within the framework of our assumptions. The rest of the mass of the Moon was assumed to accrete linearly in size during the initial 100 years. The thermal profiles and the deduced planetary scale evolutionary trends of Moon are presented in Figure 1-5. The majority of the simulations were run for the initial couple of million years.

The thermal profiles corresponding to all the Models at the end of complete accretion of Moon over an assumed timescale of 100 years are presented in Figure 1. The pressure dependent solidus and liquidus temperatures for the melting of bulk silicates and metallic iron (Fe-FeS) are presented as a function of depth in Figure 1. The inset in the figure shows the deduced pressure-depth profile of undifferentiated Moon with an assumed uniform density. The influence of the adoption of the modified relaxation method for the rapid cooling of the convective magma ocean is marked in all the simulations as the temperature stabilizes corresponding to at an assumed ~43 % bulk silicate melting even with the varied values of the impact energy efficiency parameter, 'h'. A thermal gradient is established in the convective magma ocean beyond the region that acquires the necessary temperature for the required amount of silicate melting by the impact induced energy. There is no further increase in the temperature beyond this region due to rapid convective heat losses even with an increase in the impact deposition energy on account of accretion. Hence, the deduced initial thermal profiles, irrespective of the varied accretion scenarios and initial temperature, are mostly identical except for the changes in the extent of the depth of the convective magma ocean that depend upon the parameter, 'h' and is independent of the size of the protoMoon accreted during the first year. The majority of the models presented here suggest substantially deep initial magma oceans that are even larger than 1000 km. This presumably support the suggestion for the possibility of large initial lunar magma ocean (Elkins-Tanton et al. 2011).

In order to study the cooling of the initial convective magma ocean that are produced by the adopted relaxation method as presented in the Figure 1, we selected Model J (Table 1) to study in details the subsequent thermal evolution with distinct cooling rates for the magma ocean (Figure 2). Model J was simulated with a value of 1700 K for $T_{ini.}$ and a value of 0.1 for 'h'. Four simulations with distinct values of $2\times10^3$, $2\times10^4$, $4\times10^4$ and $2\times10^5$ for $\kappa_{conv.max.}/\kappa_{cond.}$ were run by using the approach based on the Equation 14 in the last section. The thermal profiles for these simulations are presented in the Figures 2a-c. During the cooling phase of Moon, the gradients in the thermal profiles within the convective magma ocean substantially reduce over time except in the outer regions. We observe a systematic trend in the cooling rate that scales proportionally with the increase in the effective thermal diffusivity. This is essentially due to the manner in which '$\kappa\,\delta t$' appears in the numerical solution for a fixed value of $\delta r$ (Equation 3). The observed trend can be used to extrapolate the cooling rate at high values of effective thermal diffusivity that have been deduced in the last section based on the theoretical studies. The cooling rate of the magma ocean along with the planetary scale evolution of Moon is presented in Fig. 2d in terms of accretion, the generation of substantial bulk silicate melting and the formation of an iron core. The radial distance that marks the maximum change in the thermal gradients on account of convective heat losses inferred for the various thermal profiles from the Fig. 2a-c are presented as dashed lines (Fig. 2d) representing the evolutionary trends for the four distinct values of $\kappa_{conv.max.}/\kappa_{cond.}$. These evolutionary dashed lines represent the extent of the depth of the effective convective magma ocean with time. Based on the four deduced evolutionary trends, we extrapolated the evolutionary trends in the extent of the effective convective magma ocean for two distinct values of $2\times10^6$ and $2\times10^7$ for $\kappa_{conv.max.}/\kappa_{cond.}$ as black arrows (Fig. 2d). The cooling rates inferred by the bound range of these two-extreme effective thermal diffusivities indicate an extensive cooling of the lunar magma ocean over timescales of $10^3$-$10^4$ years. This seems to be compatible with the suggestion of ~80 % cooling of lunar magma ocean over an identical duration (Elkins-Tanton et al. 2011). The cooling of the remaining magma ocean can prolong for several tens of million years as indicated by the chronological records of anorthositic crust samples.

The radial growth of the iron core over time is also presented in Figure 2d. The growth of the iron core in the present work is limited primarily by the timescale over which the distinct regions of a planetary body achieve the basic assumed requirement of 40 % bulk silicate melting that is necessary to initiate metal-silicate segregation, and the viscosity dependent descent velocity of the molten iron blobs through the molten silicate mush (Taylor et al. 1993; Rushmer et al. 2000; Bhatia and Sahijpal 2017a). Due to the uncertainties in precisely estimating the descent velocities, we present a temporal range over which the core-mantle segregation could effectively take place. The solid black region at the center in Fig. 2d, represents the observed growth rate trend estimated by our simulations. It should be noted that the substantial extent of partial segregation initiates well within the initial 100 years, however, the trends presented here represent complete segregation that initiates around 140 years and is complete by ~330 years leading to the formation of a ~328 km sized iron core. The estimated trends seem to be compatible with the earlier work (see e.g., Rushmer et al. 2000). However, in case if we use the maximum descent velocity value of ~1 km year$^{-1}$ (Bhatia and Sahijpal 2017a), the core-mantle differentiation could prolong for 1000-2000 years. A typical molten metallic blob would sink from the lunar surface to the core during this time-span. The white arrows within the growing iron core (Fig. 2d.) represent the temporal range over which the core-mantle differentiation could effectively prolong. We have not considered the role of Rayleigh–Taylor instability in triggering rapid core-mantle differentiation (Srámek et al. 2012) The other physical process that can enhance the downward descent of molten metallic iron is the coalescence of molten metallic blobs (Senshu et al. 2002). We anticipate a comparatively rapid core formation if these two effects are incorporated.

It should be noted that the substantial amount of magma ocean cools down over an identical duration. The inner regions of this cooling magma ocean would eventually start solidifying due to crystallization and Ostwald ripening (Solomatov 2007). Hence, the cooling of the magma ocean could substantially lower the descent velocity of the molten metallic blobs through viscous silicate mush. This could result in a sluggish metal segregation through porous flow, with a melt percolation velocity of ~0.2 km year$^{-1}$ according to Darcy's law (Rushmer et al. 2000), thereby, leading to the growth of an iron core over a timescale of several thousands of years. We do not rule out the possibility of melt percolation through porous flow as the prime cause for core-mantle differentiation. This porous flow could have occurred even with substantially low extent of bulk silicate melting, thereby, resulting in the probability of formation of a magma ocean with smaller depths. Shallow magma oceans are also feasible with an efficient molten metallic porous flow through bulk silicates occurring over a timescale of thousands of years.

If we assume that the entire Moon accreted in a totally molten state with $T_{ini.} > 1800$ K, the need of the impact induced energy becomes almost redundant. We ran a simulation with a value of 1800 K for $T_{ini.}$ and a value of 0.1 for h. The results are not graphically presented here since this simulation is not distinct from the one that are presented here. This is essentially due to the fact that even with a high $T_{ini.}$ value, the maximum temperature achieved at the center does not exceed ~1750 K due to the adopted relaxation method. The iron core will grow over a timescale of 100-1000's years and the magma ocean will substantially cool down over $10^3$-$10^4$ years.

As discussed in the earlier section we could incorporate the dependence of the effective thermal diffusivity on the acceleration due to gravity, 'g(r)', in a normalized manner (Equation 14) and the dependence on the depth of the magma ocean could not be successfully incorporated. We observed a minor temperature dip with a maximum value of ~15 K near the center of the cooling Moon. This temperature dip eventually reduces with temporal evolution. We attribute this dip to the dependence on the effective thermal diffusivity on g(r) that shows a strong radial dependence at the center (Equation 14). The temperature dip disappears if we instead consider a linear power of g(r) rather the presently adopted power of 0.33. We have, however, retained the dependence according to Equation 14. We anticipate that the incorporation of the depth dependence of the magma ocean (Neumann et al. 2014) in Equation 14 could make the approach more robust.

The approach adopted in the case of Model J, regarding the prediction of the cooling rate of magma ocean at high effective thermal diffusivity, and the predicted growth rate of iron core is extended in the remaining simulations. Models A-D, with an assumed $T_{ini.} \leq 1600$ K infer a partial differentiated body that results in the growth of an iron-shell instead of an iron core (Fig. 3). This iron-shell represents a thin shell of partially differentiated iron that does not further descent downwards due to the assumptions made in the present work. This shell is produced on account of substantial melting of silicates in the outer regions due to impact induced heating. We do not totally rule out the possibility of melt percolation for further differentiation. An increase in the value of h from 0.1-0.5 in the case of Models B-D results in a systematic increase in the size of the iron-shell. The earlier works associated with the incorporation of impact induced energy also indicated that the impacts during accretion provided substantial energy for significant silicate melting and planetary scale differentiation in the outer regions of the planetary bodies with radius > 1500 km even with h ~ 0.1 (Bhatia and Sahijpal 2016, 2017a, b). However, the impact induced energy alone cannot explain the core-mantle differentiation unless the initial averaged temperature ($T_{ini.}$) of the accreting Moonlets is assumed to be higher than 1600 K (Figures 2-5).

The criterion for the onset of core-mantle differentiation in the present work is based on the melting of 40 % bulk silicates (Taylor et al. 1993; Rushmer et al. 2000) in majority of the simulations. This is certainly a debatable issue due to the lower abundance of metallic iron in Moon compared to other planetary bodies. The physical mechanism responsible for the segregation of molten metallic fraction from silicate mush could be significantly different for Moon. It was assumed that the molten iron blobs descent downward towards the center only if the successive regions beneath a specific region of iron blobs have more than 40 % bulk silicates in melt form (Sahijpal et al. 2007; Bhatia and Sahijpal 2016, 2017a). If we assume 10 % bulk silicate melting as the essential criteria for the initiation of metal-silicate segregation, the core-mantle differentiation can be triggered even with a value of 1600 K for the initial averaged temperature ($T_{ini.}$) as indicated in the Figure 4. Model H and Model G were run with an assumed requirement of 10 % and 40 % bulk silicate melting, respectively, for the triggering of core-mantle differentiation. The differentiation occurs in the former case, whereas, only a thin iron-shell is produced in the latter case. Model H yields a comparatively small magma ocean depth of ~1000 km compared to the other models with a higher value of $T_{ini.}$ that produce core-mantle differentiation (Figures 2-5).

In majority of our simulations the depth of the initial convective magma ocean is larger than the one anticipated in some of the earlier studies (Warren 1985; Wieczorek et al. 2013) except if we consider substantial contribution from the melt percolation through porous flow, as discussed earlier. However, our results, specifically in the case of Model H, seems to be compatible with the magma ocean depths suggested by Elkins-Tanton et al. (2011). If we further relax the condition for metal-silicate segregation at a lesser level of bulk silicate melting, the depth of the magma ocean will further reduce. Nonetheless, this scenario will still produce core-mantle differentiation. We anticipate a collective role of both melt percolation through porous flow as well as viscosity dependent descent of molten metallic blobs through silicate melt (Rushmer et al. 2000).

On account of impact induced heating in case the outer regions of the Moon experience extensive heating (> 2500 K), the vaporization of silicates and loss of the accreted matter could occur from the surface. The loss is replenished by the accretion of matter from the protolunar disk. The replenished matter could include contributions from the vaporized silicates that re-condense and accrete back on Moon. The environment prevailing in the protolunar disk can support the possibility of multiple episodes of accretion and vaporization leading eventually to the formation of a Moon in suitably molten state to cause core-mantle differentiation.

The models with distinct initial seed radii for the protoMoon accreted during the initial one year infer almost identical results (Table 1, Figures 3-5). Further, we do not infer major differences in terms of core-mantle differentiation rate and the cooling rates of magma ocean among the simulations with the variation of the parameter, 'h' in the range of 0.1-0.2 (Figure 5). The majority of these simulations indicate the core-mantle differentiation over a timescale of 100-1000's years, whereas, substantial cooling of the convective magma ocean occurs during the initial $\sim 10^3$-$10^4$ years.

**Conclusions**

The dependence of the planetary differentiation of Moon on some of the critical parameter associated with the rapid accretion of Moon was analyzed based on the numerical simulations. These parameters include, a) the efficiency of the impact induced heating against heat losses from the planetary surface, b) the initial average temperature of the accreting Moonlets, c) the size of the protoMoon that accreted rapidly beyond the Roche limit within the first year after the giant impact, d) the contributions of the long-lived nuclides in the global heating of Moon. We have theoretically estimated the cooling rates of the convective magma ocean. The convective cooling was numerically executed by initially relaxing the thermal gradient in the accreting Moon by approximating with the gradient with an assigned melting temperature of the bulk silicates. Subsequent to the complete accretion, the cooling was parametrically performed by adopting a theoretical effective thermal diffusivity. The important findings of the present numerical simulations are;

i) The core-mantle differentiation in Moon could have occurred over a timescale of 100-1000's years. The substantial cooling of the convective magma ocean could have occurred over timescale of $10^3$-$10^4$ years.

ii) In order to heat the inner regions for differentiation, the initial average temperature of the accreting Moonlets should be high ($\geq$ 1600 K). This could be possible as the accretion of Moonlets happens in an energetic environment involving repeated episodes of vaporization and re-condensation of silicate grains.

iii) The initial size of the protoMoon accreted beyond the Roche limit within the first year does not significantly influence the thermal evolution of Moon.

iv) The contribution of the long-lived nuclides in globally heating the Moon is not significant due to low potassium.

v) The majority of our models indicate substantially large convective magma oceans with depths more than 1000 km. This constraint comes from the commencement of core-mantle differentiation at ~40 % bulk silicate melting. However, this condition can be relaxed if the requirement of a higher fraction of the bulk silicate in the molten form could be reduced. Alternatively, the melt percolation through porous flow does not require rigorous bulk silicate melting. In such a scenario, it is possible to produce magma oceans of comparatively smaller depths.

**Acknowledgements:** We are extremely grateful to the numerous comments and suggestions made by two reviewers and the associate editor. This work was supported by a PLANEX (ISRO) research grant.

**Table 1.** Simulation parameters for the early thermal evolution of Moon

| S.No. | Models | Seed radius (km) | Initial temperature ($T_{ini.}$ K) | h | Figure |
|---|---|---|---|---|---|
| 1 | Model A | 1074 | 1400 | 0.5 | Fig. 3a, b |
| 2 | Model B | 1074 | 1600 | 0.1 | Fig. 3c, d |
| 3 | Model C | 1074 | 1600 | 0.2 | Fig. 3e, f |
| 4 | Model D | 1074 | 1600 | 0.5 | Fig. 3g, h |
| 5 | Model E | 1074 | 1700 | 0.1 | Fig. 4a, b |
| 6 | Model F | 1074 | 1700 | 0.2 | Fig. 4c, d |
| 7 | Model G | 1240 | 1600 | 0.1 | Fig. 4e, f |
| 8 | Model H | 1240 | 1600 | 0.1 | Fig. 4g, h |
| 9 | Model I | 1240 | 1600 | 0.2 | Fig. 5a, b |
| 10 | Model J | 1240 | 1700 | 0.1 | Fig. 2a-d |
| 11 | Model K | 1240 | 1700 | 0.15 | Fig. 5c, d |
| 12 | Model L | 1240 | 1700 | 0.2 | Fig. 5e, f |

$T_{Onset}$ = 30 to >100 Million years in the early solar system; $T_{acc}$ = 100 years is the assumed accretion duration of Moon. The seed radius defines the size of the protoMoon that accretes beyond the Roche limit during the first year. The temporal and spatial grid sizes were assumed to be 0.1 years and 2 km, respectively, in all the simulations except for one of the simulations for the Model J (Fig. 2c). The simulations were run up to the initial couple of million years in most cases. Except for the Model H, the core-mantle differentiation in the remaining models was triggered subsequent to 40% bulk silicate melting. In case of Model H, the differentiation was triggered at 10% bulk silicate melting.

The initial abundances of the long-lived nuclides, $^{40}$K (τ ~ 1.82 Gyr), $^{235}$U (τ ~ 1 Gyr), $^{238}$U (τ ~ 6.5 Gyr), and $^{232}$Th (τ ~ 20 Gyr), with the decay energies of ~0.71 MeV, 45.9 MeV, 48.1 MeV and 40.44 MeV, respectively, for the volatile-depleted Moon were modified from the estimates of the initial solar system abundances presented by Lodders et al. (2009). The major change occurs due to the significant reduction in volatile $^{39}$K from a value of 0.07 wt. % to 0.0037 wt. %, with the initial $^{40}$K/$^{39}$K = 1.58 × 10$^{-3}$.

**Figure Captions**

**Fig. 1.** The deduced thermal-depth profiles for different accretion scenarios after the complete accretion of Moon (Table 1). The accretion of Moon was commenced within 100 years after the giant impact of the impactor proto-planet with the early Earth around 30-100 million years ($T_{Onset}$) in the early solar system. The pressure dependent solidus and liquidus phase transition temperatures for the bulk silicates and metallic iron are represented at various depths. The impact energy efficiency parameter (h) is defined according to the Equation 5. The deduced pressure-depth dependence of the undifferentiated Moon with an assumed uniform density is shown in the inset figure.

**Fig. 2a, b, c.** The deduced thermal evolution of Moon for the selected model J (Table 1) with a value of 1700 K for $T_{ini.}$ (an initial average accretion temperature of Moonlets) and a value of 0.1 for h, for four distinct values of $2\times10^3$, $2\times10^4$, $4\times10^4$ and $2\times10^5$ for $\kappa_{conv.max.} / \kappa_{cond.}$. The solid and the dashed line curves in the Fig. 2b correspond the lower and higher values of $\kappa_{conv.max.} / \kappa_{cond.}$, respectively. The planetary scale evolution of Moon is presented in Fig. 2d in terms of accretion, the bulk silicate melting (marked in green color) and the formation of an iron core (marked in black color). The dashed lines represent the evolutionary trends for the four distinct values of $\kappa_{conv.max.} / \kappa_{cond.}$ in terms of the spatial region over which maximum change in the thermal gradient on account of convective heat losses occurs. The extent of the effective convective magma ocean was deduced from the three distinct thermal profiles in Fig. 2a, b, c. The extrapolated evolutionary trends in the extent of the effective convective magma ocean for two distinct values of $2\times10^6$ and $2\times10^7$ for $\kappa_{conv.max.} / \kappa_{cond.}$ is presented in Fig. 2d as black arrows. We anticipate a thermal evolution well within the extreme bounds of these two evolutionary trends. The white arrows within the growing iron core represent the temporal range over which the core-mantle differentiation could effectively prolong. The melt percolation through porous flow can further prolong the timescale by a factor of four if we assume a melt percolation velocity of ~0.2 km year$^{-1}$.

**Fig. 3a, c, e, g.** The deduced thermal evolution of Moon for the distinct simulation parameters (Table 1) for a value of $2\times10^4$ for $\kappa_{conv.max.} / \kappa_{cond.}$. The corresponding planetary scale evolution of Moon is presented in **Fig. 3b, d, f, h**, in terms of accretion, the generation of substantial bulk silicate melting (marked in green color) and the formation of an iron-shell/iron core (marked in black color). The un-melted regions are marked in yellow colors. The evolutionary trends in the extent of the effective convective magma ocean for two distinct values of $2\times10^6$ and $2\times10^7$ for $\kappa_{conv.max.}$ / $\kappa_{cond.}$ is presented in the figures as black arrows. We anticipate a thermal evolution well within the bounds of these two evolutionary trends. The white arrows within the growing iron core represent the temporal range over which the core-mantle differentiation could effectively prolong. The melt percolation through porous flow can further prolong the timescale by a factor of four.

**Fig. 4.** Identical to Fig. 3, except for distinct simulation parameters.

**Fig. 5.** Identical to Fig. 3, except for distinct simulation parameters.

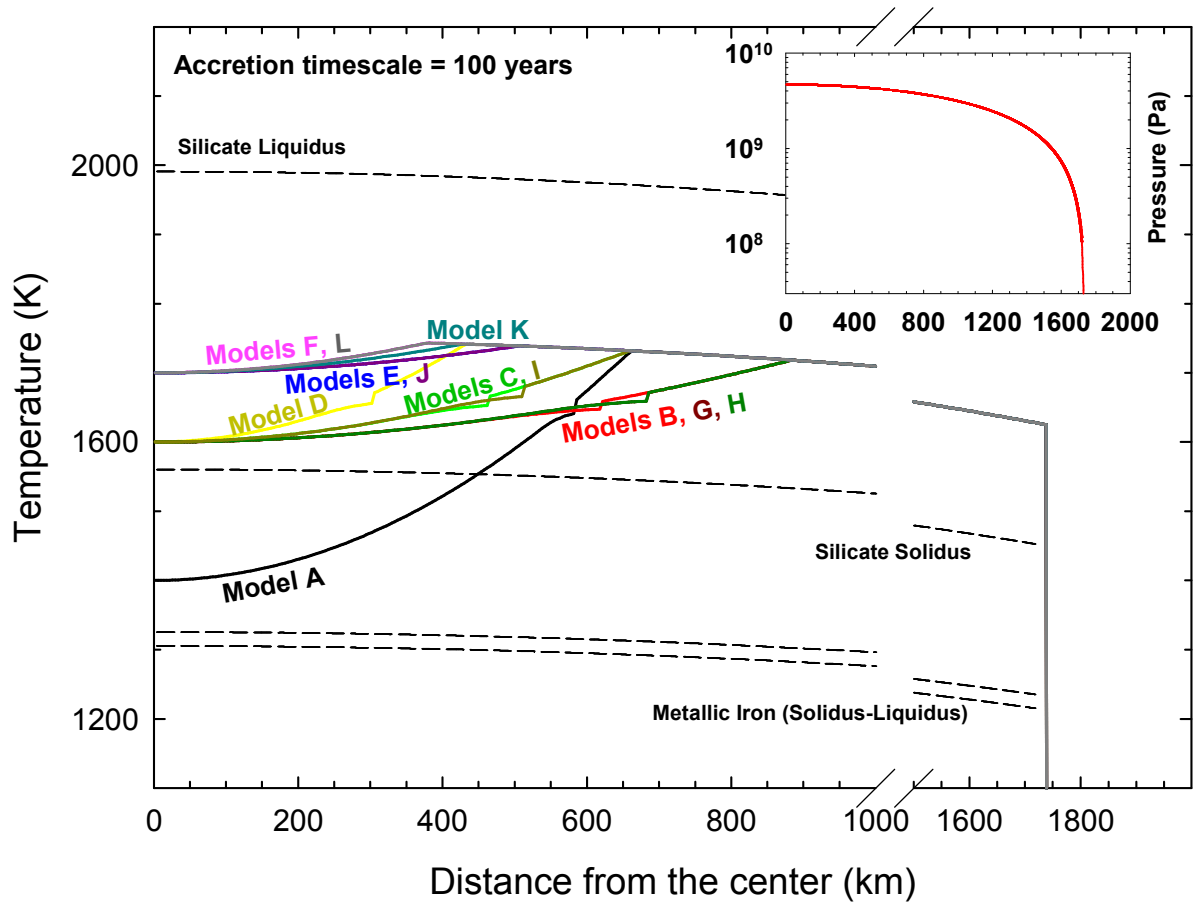

Figure 1

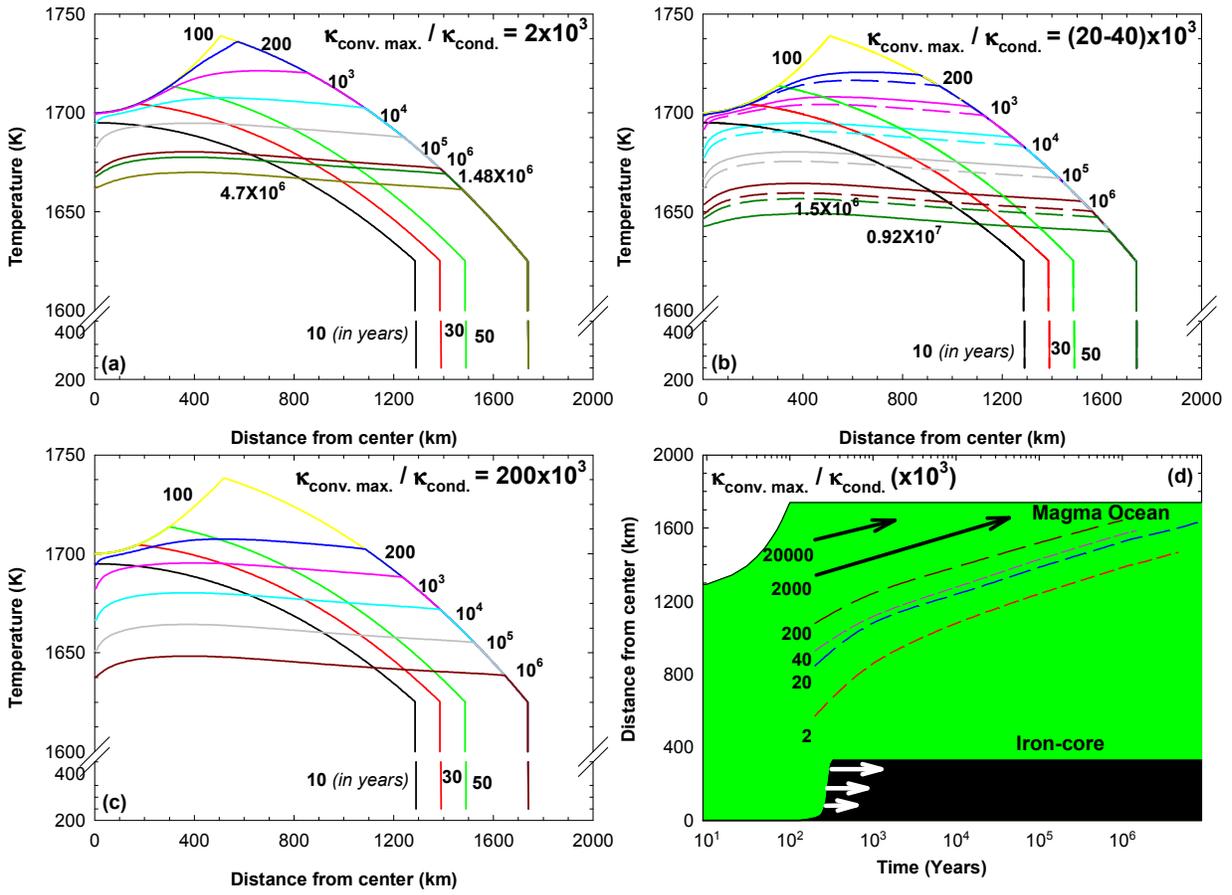

**Figure 2**

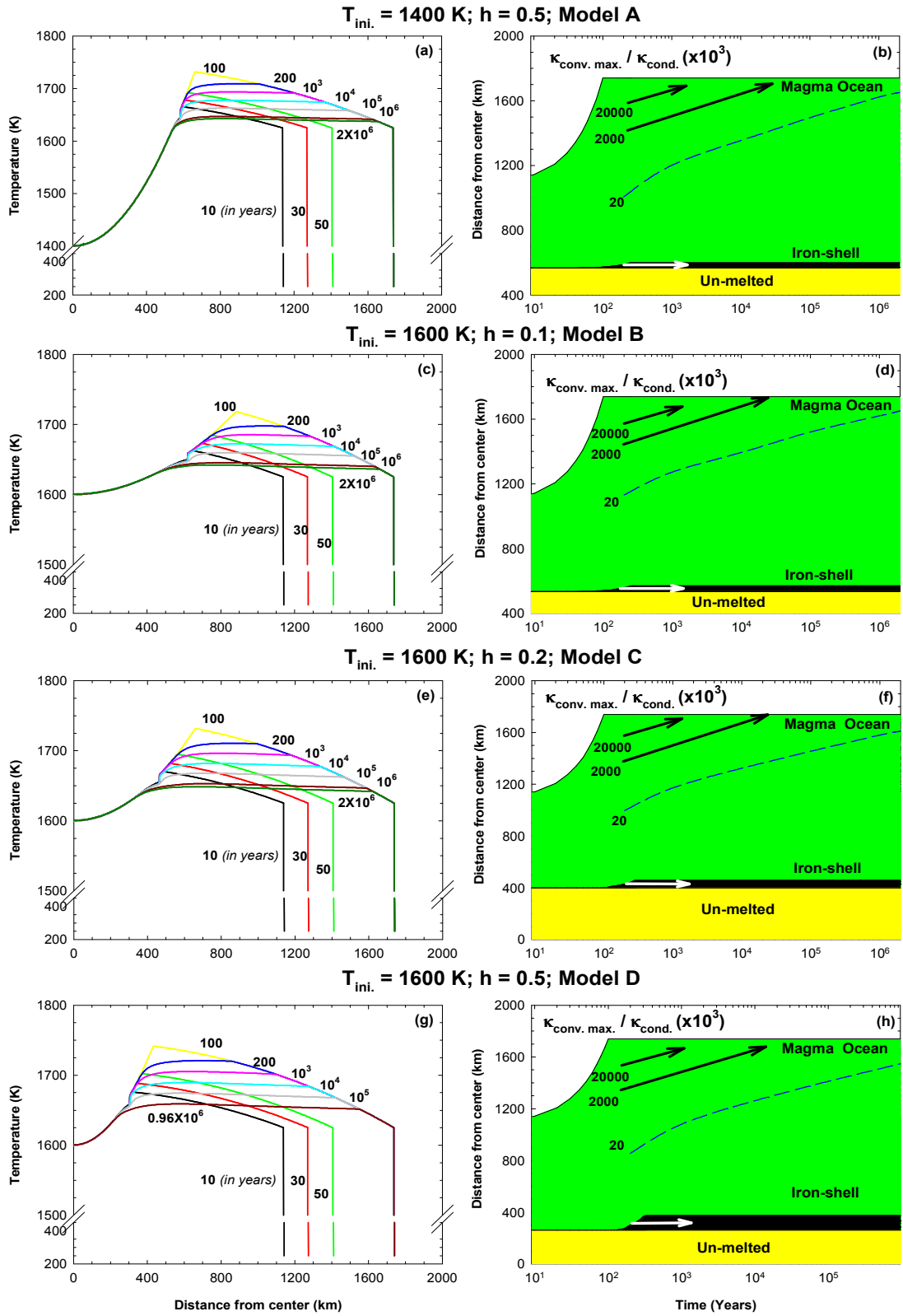

**Figure 3**

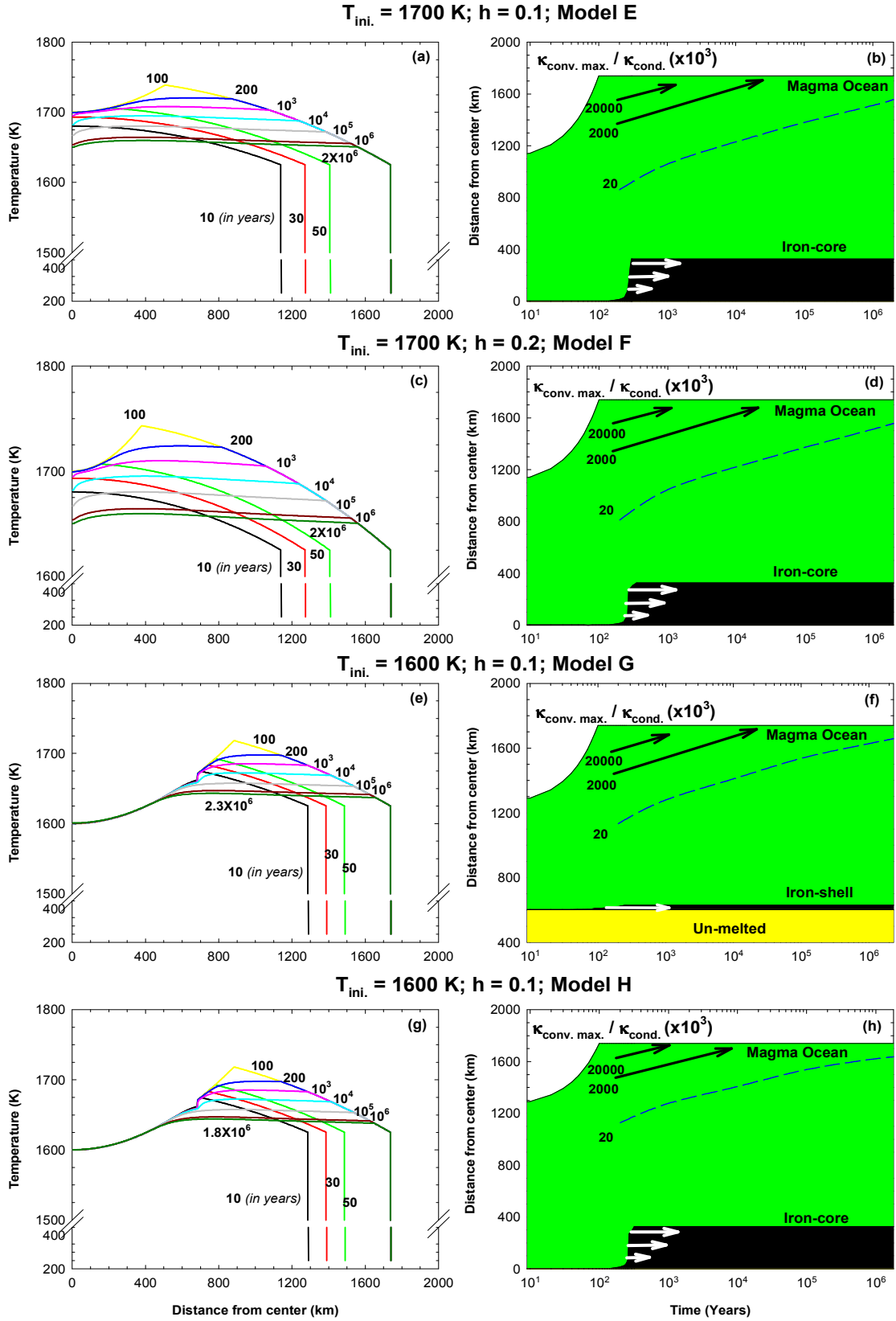

Figure 4

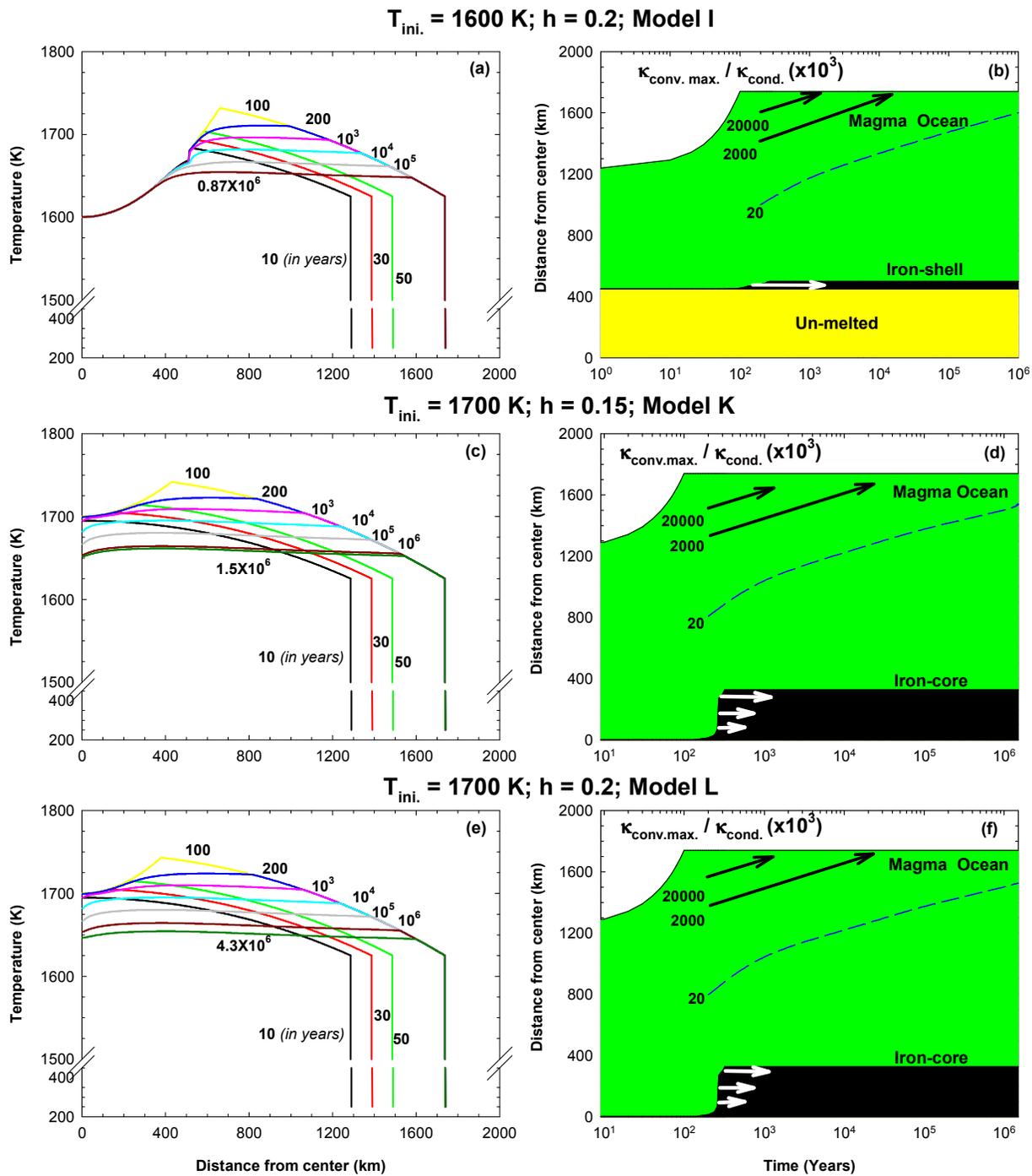

Figure 5